\documentclass[journal, comsoc]{IEEEtran}
\usepackage{amsmath,amsfonts}
\usepackage{algorithmic}
\usepackage{array}
\usepackage{textcomp}
\usepackage{stfloats}
\usepackage{url}
\usepackage{verbatim}
\usepackage{graphicx}
\usepackage{acro}
\usepackage{cite}
\usepackage{braket}
\usepackage{subcaption}

\usepackage{tikz}
\usepackage{siunitx}

\newcommand{\ok}{\checkmark}
\newcommand{\ko}{\mathcal{X}}

\DeclareMathOperator{\dB}{dB}

\DeclareAcronym{AO}{short = AO , long = Adaptive Optics}
\DeclareAcronym{BB84}{short = BB84 , long = Bennett and Brassard{,} 1984}
\DeclareAcronym{BF}{short = BF, long = Bit Flipping}
\DeclareAcronym{BP}{short = BP, long = Belief Propagation}
\DeclareAcronym{BSC}{short = BSC , long = Binary Symmetric Channel}
\DeclareAcronym{CV-QKD}{short = CV-QKD , long = Continuous Variable \ac{QKD}}
\DeclareAcronym{DLL}{short = DLL , long = Dynamic-Link Library}
\DeclareAcronym{DLR}{short = DLR , long = Deutsches Zentrum f\"ur Luft- und Raumfahrt}
\DeclareAcronym{DV-QKD}{short = DV-QKD , long = Discrete Variable \ac{QKD}}
\DeclareAcronym{EB}{short = EB , long = Entanglement-Based}
\DeclareAcronym{EC}{short = EC , long = Error Correction}
\DeclareAcronym{ECC}{short = ECC , long = Elliptic Curve Cryptosystems}
\DeclareAcronym{FSO}{short = FSO , long = Free Space Optical}
\DeclareAcronym{GNFS}{short = GNFS , long = General Number Field Sieve}
\DeclareAcronym{HV}{short = HV , long = Hufnagel-Valley}
\DeclareAcronym{IR}{short = IR , long = Information Reconciliation}
\DeclareAcronym{LDPC}{short = LDPC , long = Low-Density Parity-Check}
\DeclareAcronym{LEO}{short = LEO , long = Low Earth Orbit}
\DeclareAcronym{LLR}{short = LLR , long = Log-Likelihood Ratio}
\DeclareAcronym{MDI}{short = MDI , long = Measurement-Device-Independent}
\DeclareAcronym{NIST}{short = NIST , long = National Institute of Standards and Technology}
\DeclareAcronym{OGS}{short = OGS , long = Optical Ground Station}
\DeclareAcronym{OGS-KN}{short = OGS-KN , long = OGS at the DLR Institute of Communications and Navigation}
\DeclareAcronym{OTP}{short = OTP , long = One Time Pad}
\DeclareAcronym{PDF}{short = PDF , long = Probability Density Function}
\DeclareAcronym{PDS}{short = PDS , long = Power Density Spectrum}
\DeclareAcronym{PM}{short = PM , long = Prepare and Measure}
\DeclareAcronym{PNS}{short = PNS , long = Photon Number Splitting}
\DeclareAcronym{PSI}{short = PSI , long = Power Scintillation Index}
\DeclareAcronym{PQC}{short = PQC , long = Post Quantum Cryptography}
\DeclareAcronym{QBER}{short = QBER , long = Quantum Bit Error Rate}
\DeclareAcronym{QKD}{short = QKD , long = Quantum Key Distribution}
\DeclareAcronym{RSA}{short = RSA , long = Rivest-Shamir-Adleman}
\DeclareAcronym{Sat-QKD}{short = Sat-QKD , long = Satellite-based \ac{QKD}}
\DeclareAcronym{SKL}{short = SKL , long = Secret Key Length}
\DeclareAcronym{SKR}{short = SKR , long = Secret Key Rate}
\DeclareAcronym{SMF}{short = SMF , long = Single-Mode Fiber}
\DeclareAcronym{SNSPD}{short = SNSPD , long = Superconducting Nanowire Single-Photon Detector}
\DeclareAcronym{SPAD}{short = SPAD , long = Single-Photon Avalanche Detector}
\DeclareAcronym{WCP}{short = WCP , long = Weak Coherent Pulses}
\DeclareAcronym{wlog}{short = wlog , long = without loss of generality}

\def\BibTeX{{\rm B\kern-.05em{\sc i\kern-.025em b}\kern-.08em
    T\kern-.1667em\lower.7ex\hbox{E}\kern-.125emX}}
\usepackage{balance}
\begin{document}

\title{Optimization of Information Reconciliation for Decoy-State Quantum Key Distribution over a Satellite Downlink Channel}

\author{Thomas Scarinzi, Davide Orsucci, Marco Ferrari,~\IEEEmembership{Member,~IEEE}, and Luca Barletta,~\IEEEmembership{Member,~IEEE}
\thanks{T. Scarinzi was with Politecnico di Milano, 20133 Milan, Italy (email: thomasscarinzi@gmail.com)}
\thanks{D. Orsucci is with the Institute of Communications and Navigation, Deutsches Zentrum f\"ur Luft- und Raumfahrt (DLR), 82234 Weßling, Germany (email: davide.orsucci@dlr.de).}
\thanks{M. Ferrari is with the Consiglio Nazionale delle Ricerche, Istituto di Elettronica e di Ingegneria dell’Informazione e delle Telecomunicazioni, Italy (email: marcopietro.ferrari@cnr.it)}
\thanks{L. Barletta is with the Dipartimento di Elettronica, Informazione e Bioingegneria, Politecnico di Milano, 20133 Milan, Italy (email: luca.barletta@polimi.it)}
}


\maketitle

\begin{abstract}
\ac{QKD} is a cryptographic solution that leverages the properties of quantum mechanics to be resistant and secure even against an attacker with unlimited computational power. Satellite-based links are important in \ac{QKD} because they can reach distances that the best fiber systems cannot. However, links between satellites in \ac{LEO} and ground stations have a duration of only a few minutes, resulting in the generation of a small amount of secure keys. In this context, we investigate the optimization of the information reconciliation step of the \ac{QKD} post-processing in order to generate as much secure key as possible. As a first step, we build an accurate model of the downlink signal and \ac{QBER} during a complete satellite pass, which are time-varying due to three effects: (i) the varying link geometry over time, (ii) the scintillation effect, and (iii) the different signal intensities adopted in the Decoy-State protocol. Leveraging the a-priori information on the instantaneous QBER, we improve the efficiency of \ac{IR} (i.e., the error correction phase) in the Decoy-State \acs{BB84} protocol, resulting in a secure key that is almost 3\% longer for realistic scenarios. 
\end{abstract}


\section{Introduction}

\IEEEPARstart{T}{he} future advent of quantum computers represents an existential security threat for the cryptographic systems that are used nowadays. When Peter Shor invented the quantum algorithm capable of factoring numbers with polynomial complexity~\cite{shor}, it basically rendered \ac{RSA} insecure against an adversary having access to a quantum computer. Additionally, the period finding primitive present in Shor's algorithm allows an attacker to break also \ac{ECC}~\cite{shor_ec}.
Furthermore, the problem is already pressing today because of the so called ``harvest now, decrypt later" attack, in which the adversaries could store the communications now and decrypt them in the future when they will have the sufficient computational power.
To address this problem, one of the possible solutions is
\ac{QKD}, which allows to securely exchange two identical keys (or passwords) between two users. These keys may later be used to secure the communication between the two users through a symmetric cryptosystem.

The development of \ac{QKD} technology for ground-based systems has reached a rather high level of maturity and is already commercially available as a certified product~\cite{certified_QKD}. A major hurdle is that when a quantum state propagates through an optical fiber, both photon loss (attenuation) and decoherence processes cause its fidelity to decay exponentially with distance, setting a hard limit on the maximum distance in which ground-based \ac{QKD} is feasible. \acs{Sat-QKD} constitutes a promising solution to this problem, introducing satellites into the game in order to connect distant users. Indeed, in free-space channels, transmission generally decays quadratically with distance due to beam divergence, with additional losses arising due to atmospheric absorption and receiver terminal finite optical transmission. This quadratic scaling enables establishing QKD links over long distances, sufficient to connect an \ac{OGS} and a satellite in \ac{LEO} with optical terminals of relatively small size. The advantage of \acs{Sat-QKD}, with respect to fiber-based systems, is that to achieve these long distance communications, only one satellite is needed as a trusted node, provided that during its orbit it can establish connections with all the users~\cite{orsucci}.
Instead, to reach these distances with ground-based systems, many relays are needed and therefore many trusted nodes on the ground, which pose more vulnerability issues compared to a single satellite in \ac{LEO}.

\ac{Sat-QKD} faces several challenges, such as weather dependence, link availability, hardware and deployment complexity, and costs. Anyway, in recent years there has been significant progress, with several missions demonstrating feasibility and operational capabilities. The Chinese satellite \textit{Micius} pioneered long-distance \ac{QKD} with successful Decoy-State \ac{BB84} key exchange over distances exceeding 1,200 km~\cite{micius}, together with the first \ac{QKD} encrypted video call between Vienna and Beijing in 2017~\cite{quantum-call}. Following this, the Jinan-1 satellite extended QKD applications in urban-scale networks and emphasized integration with terrestrial fiber infrastructures~\cite{jinan-1}, while Tiangong-2 demonstrated \ac{QKD} with small-sized payload. In Europe, missions such as \textit{SAGA} and \textit{Eagle-1} are designed to deploy \ac{PM} \ac{QKD} payloads in \ac{LEO}, aiming for operational secure links between ground stations~\cite{saga, eagle-1}. More recently, CubeSat missions such as \textit{QUBE}, developed by the \ac{DLR}, and its successor \textit{QUBE-II}, aim to validate miniaturized quantum payloads and investigate their performance in constrained orbital environments~\cite{qube,qube2}. These efforts collectively indicate a shift from experimental demonstrations to the early stages of operational satellite \ac{QKD} constellations. In this work, the objective is to optimize one of the most commonly used protocols, that is the Decoy-State \ac{BB84}, in the context of \ac{Sat-QKD}, by using the instant-by-instant channel model, instead of using the average loss during the downlink communication. The main factors in play are the transmission change due to the changing link distance and the turbulence-induced scintillation.

\paragraph*{Paper novelty} 
The main  contribution of this work is the optimization of the \ac{IR} phase for Decoy-State BB84 under non-constant  error rates. Such conditions arise in satellite-to-ground \ac{QKD} downlinks due to channel dynamics, where the use of multiple pulse intensities in the Decoy-State protocol induces a signal-dependent \ac{QBER}. In practice, the \ac{QBER} can be accurately estimated by monitoring  channel transmission with optical support equipment already required for beaconing and  classical communication, as well as  by measuring the error rate in the non-key-generating basis. Exploiting this a-priori information enables  more efficient error correction during the \ac{IR} stage of \ac{QKD} post-processing, thereby extending the secure key length without hardware modifications and with only minor changes to the post-processing.

\paragraph*{Paper organization} 
The remainder of this paper is organized as follows. Sec.~\ref{sec:preliminaries} introduces the background on \acl{IR} and \ac{QKD}. Sec.~\ref{sec:sat_model} covers the link budget and Sec.~\ref{sec:scintillation} the power scintillation, which combined give the downlink channel model. Sec.~\ref{sec:QKD_model} present the model of the \ac{QBER} in the \ac{QKD} receiver. In Sec.~\ref{sec:methodology} we present details on the simulation implementation, whose result are presented in Sec.~\ref{sec:simulations}, evaluating the performance of adaptive \ac{LDPC}-based \ac{IR} schemes under realistic channel conditions. Finally, Sec.~\ref{sec:conclusions} concludes the paper and discusses directions for future work.

\section{Preliminaries}
\label{sec:preliminaries}
This section provides the necessary background to understand the design and optimization of \ac{IR} for \ac{Sat-QKD}. 

\subsection{Information Reconciliation with LDPC Codes}
Consider the case where Alice and Bob have two unequal but correlated bit strings, $\mathbf{z}_A,\ \mathbf{z}_B$, obtained by \textit{sifting} a larger sequence of prepared and measured states. In direct \acl{IR} Bob tries to retrieve the value $\mathbf{z}_A$ while exchanging as little information as possible, similarly to what is done in error correction. The Cascade protocol~\cite{cascade} and \ac{LDPC}-based schemes are among the most widely adopted approaches for \ac{IR}. Here we focus on the latter, which were shown to have good performance for \ac{IR} in \ac{QKD} scenarios~\cite{protograph}.

Let $H$ be the parity check matrix of a (fixed) \ac{LDPC} code. Alice computes the syndrome vector $\mathbf{s}_A = H \cdot \mathbf{z}_A \pmod 2$ and sends the massage to Bob over a public channel. Bob locally computes $\mathbf{s}_B = H \cdot \mathbf{z}_B \pmod 2$ and together with $\textbf{s}_A$ he gets
\begin{align}
    \mathbf{s} = \mathbf{s}_A + \mathbf{s}_B =
    H \cdot (\mathbf{z}_A + \mathbf{z}_B) = H \cdot \mathbf{e}
\end{align}
which means that $\mathbf{s}$ is the syndrome associated with the error vector $\mathbf{e}$. If the syndrome $\mathbf{s}$ has sufficient information on $\mathbf{e}$, a decoder can succeed in recovering its value.

In this work we consider \ac{IR} based on Rateless Protograph \ac{LDPC} codes. The protograph is a small bipartite graph that is used to generate the Tanner graph of the LDPC code by replicating the protograph multiple times and then connecting the copies by applying some circular shifts~\cite{ldpc_from_protograph}.  The main advantage of protograph codes for \ac{IR}  is that a single protograph can be used to generate \ac{LDPC} codes of different block lengths by varying the lift factor (i.e. the number of replications of the protograph), thereby providing flexibility for different message lengths without requiring the design of a new code in each case. Besides, the convergence properties of BP decoding can be accurately predicted based on the protograph structure. In a rateless implementation, a sequence of parity-check matrices with decreasing code rates is defined a priori. For each reconciliation attempt, the most suitable code is selected, prioritizing the one with the lowest inefficiency; if decoding fails, the code rate is progressively lowered until convergence is achieved. 
In \ac{IR}, the empirical inefficiency factor $f$ quantifies the multiplicative overhead in information leakage relative to the asymptotically achievable limit. It is defined as
\begin{align}
    f = \frac{1-R}{h_2(\phi)},
\end{align}
where $\phi$ the error rate and $R$ is the code rate at which convergence is reached, given by
\begin{align}
    R = \frac{n-m}{n}
    \label{eq:rate}
\end{align}
with $m$ the length of the syndrome $\textbf{s}$ and $n$ the length of the sifted key. The \ac{IR} rate is a critical parameter in \ac{QKD}, as it directly impacts the final \ac{SKL}: for every information bit revealed over the public channel, Alice and Bob must remove one bit from the corrected key to eliminate any knowledge potentially gained by Eve.

With rateless codes, the rate is automatically adapted to the channel reliability, without requiring an accurate a priori estimate of the error rate, thereby reducing  inefficiency. Notably, this procedure converges within only a few iterations (almost always fewer than three), whereas Cascade typically requires hundreds of iterations~\cite{cascade}. This difference is especially critical in \ac{Sat-QKD}, where the long link distance leads to round-trip times of on the order of tens of milliseconds.

\subsection{Fundamentals of Quantum Key Distribution}
\label{subsec:qkd}
\acs{QKD} is the branch of quantum cryptography devoted to the distribution of secure keys between two endpoints with information-theoretic security, thereby providing a fundamental solution to the vulnerability introduced by Shor's algorithm. 
In this work, we focus on BB84, the earliest and simplest \acs{QKD} protocol, proposed by Bennett and Brassard in 1984~\cite{BB84}. 

For concreteness, we consider qubits  encoded in the  polarization degree of freedom.
Conventionally, the $Z$-basis vectors are associated with the linear polarization states $\ket{H} = \ket{0}$ and $\ket{V} = \ket{1}$, where $H$ and $V$ denote  horizontal and vertical polarization, respectively. The $X$-basis vectors correspond to diagonal $(\ket{D} = \ket{+})$ and anti-diagonal $(\ket{A} = \ket{-})$ polarizations. Since both bases are dichotomic, each can encode a single bit of information. After the quantum transmission, Alice and Bob discard all outcomes where they used different bases (e.g., Alice used $X$ while Bob used $Z$). This step, called \emph{sifting}, yields the \emph{sifted key}. Typically, one basis (e.g., $Z$) is used for key generation, while the other (e.g., $X$) is reserved for parameter estimation.

To optimize protocol performance, it is beneficial to increase the probability $p_Z$ of selecting the key-generation basis ($Z$ basis). The probability that both Alice and Bob choose this basis is then $p_Z^2$, while the probability that both choose the parameter-estimation basis ($X$ basis) is $(1-p_Z)^2$. Hence, basis-selection  probabilities become an optimization parameter of the protocol, whereas in the original BB84 formulation the $Z$ and $X$ bases were chosen with equal probability.

In essence, the security of BB84 relies on Eve's lack of knowledge about Alice's encoding basis. If she measures qubits in randomly chosen bases, she inevitably introduces errors, which increase the \ac{QBER}. 

The original BB84 protocol assumes perfect single-photon sources, which remain experimentally challenging. Instead, we consider an implementation using attenuated laser pulses, such that the mean photon number per pulse is less than one~\cite{reviewQKD}. These \ac{WCP} introduce a vulnerability to \ac{PNS} attacks: whenever multiple photons are emitted, Eve could store one in a quantum memory  and measure it later in the correct basis, without introducing detectable errors. A countermeasure, proposed in~\cite{decoy}, is the use of  \emph{decoy states}. In this approach, Alice randomly modulates each pulse  among two or more mean intensity levels. After Bob's detection stage, Alice reveals the chosen intensities, enabling a statistical test that guarantees  a lower bound on the number of single-photon events  received by Bob. On this subset of events, the standard BB84 security proof applies.

Using the Decoy-State method, Alice and Bob employ the observed measurement outcomes to derive lower bounds on the number of vacuum events ($s_{Z,0}^-$) and single-photon events ($s_{Z,1}^-$) in the $Z$ basis, as well as an upper bound on the virtual \ac{QBER} in the $X$ basis for single-photon events that were actually measured in the $Z$ basis ($\phi_{X,1}^+$). These bounds are then used to estimate the length $\ell$ of the extractable secret key~\cite{security_bounds, consolidated_security_proof} 
\begin{align}
\label{eq:SKL}
\ell = s_{Z,0}^- + s_{Z,1}^- \bigl(1 - h(\phi_{X,1}^+)\bigr) 
       - \lambda_\text{IR} 
       - \log_{2}\frac{2}{\varepsilon_{\text{cor}}}
       - 6 \log_{2}\frac{21}{\varepsilon_{\text{sec}}}
\end{align}
where $\lambda_\text{IR}$ is the information leaked during \ac{IR}, while $\varepsilon_{\text{cor}}$ and $\varepsilon_{\text{sec}}$ are the correctness and security parameters of the protocol, respectively. Specifically, $\varepsilon_{\text{cor}}$ is the maximum probability that Alice's and Bob's keys differ, and $\varepsilon_{\text{sec}}$ is the maximum probability that any information about the key is available to Eve. In this work, we set $\varepsilon_{\text{cor}} = 10^{-15}$ and $\varepsilon_{\text{sec}} = 10^{-9}$.

\subsection{Average vs Instantaneous Quantum Bit Error Rate}

The empirical average \ac{QBER} is defined as the ratio between the total number of errors and the total number of established bits. 
\begin{align}
    \phi_X & \equiv \text{QBER}_X :=
    \frac{|\mathbf{x}_A \oplus \mathbf{x}_B|}{\text{len}(\mathbf{x}_A)}, \\
    \phi_Z & \equiv \text{QBER}_Z :=
    \frac{|\mathbf{z}_A \oplus \mathbf{z}_B|}{\text{len}(\mathbf{z}_A)},
\end{align}
where $\mathbf{x}_A$ and $\mathbf{x}_B$ (respectively, $\mathbf{z}_A$ and $\mathbf{z}_B$) denote the binary vectors representing the sifted bits of Alice and Bob in the $X$ (resp.~$Z$) basis. Here, $|\mathbf{x}|$ indicates the Hamming weight of $\mathbf{x}$ (i.e.,  the number of non-zero elements). 

In~\cite{devetak-winter}, it was shown that the asymptotic key rate of a \ac{QKD} protocol can be expressed as
\begin{align}
    R = 1 - h_2(\phi_X) - h_2(\phi_Z),
\end{align}
where $h_2(x) := - x \log_2(x) - (1-x) \log_2(1-x)$ is the binary entropy function. In this expression, 
$h_2(\phi_X)$ represents an upper bound on the information available to Eve, while $h_2(\phi_Z)$ denotes the asymptotic lower bound on the  information required for information reconciliation. In particular, when $\text{QBER}_X = \text{QBER}_Z = \phi$, the threshold \acs{QBER} above which the key rate vanishes is obtained by solving
\begin{align}
    1-2h_2(\phi) \geq 0,
\end{align}
which yields $\phi \lesssim 11\%$.

When the channel is  time-varying, each qubit (or each \ac{WCP}) can experience a different error probability. This is the case in \ac{FSO} satellite channels, where Alice and Bob may  estimate the instantaneous error probability. For instance, classical optical support systems (e.g., classical communication or beacon signals) can be used to infer the probability that a single photon transmitted by Alice through the quantum channel is detected by Bob, since classical and quantum channels exhibit proportional transmission. This enables instantaneous error-rate estimation: for example, if detector dark counts and background light dominate and remain constant (or vary slowly), the error rate increases whenever the quantum channel transmission decreases.

This a priori reliability information can be exploited to improve the efficiency of \ac{IR}. One possible strategy is to group  bits into blocks with approximately equal error probability $\phi_j$, and to perform \ac{IR} independently on each block. Asymptotically, the average information required for \ac{IR} is
\begin{align}
    \lambda_\text{IR} \approx \sum_{j} n_j h_2(\phi_j) \le n h_2(\langle\phi\rangle), 
\end{align}
where $n_j$ is the number of bits with error probability $\phi_j$, and $\langle\phi\rangle = \frac{1}{n} \sum_j n_j \phi_j$ is the average \ac{QBER}. The inequality follows from the concavity of the binary entropy. This approach allows extraction of a longer secure key. However, due to finite-size effects, this advantage emerges only for very large block sizes. 

A more practical strategy is to apply \ac{IR}  to large blocks of bits with different reliabilities, while providing the reliability estimates to the LDPC decoder in the form of \ac{LLR}s. An efficient decoder can then exploit this side information to approach the asymptotic rate.

The objective of the present work is to exploit such a priori reliability information  to improve the efficiency of \ac{IR} in Decoy-State BB84. This reduces the  information leakage, $\lambda_\text{IR}$, thereby increasing the secret key length according to~\eqref{eq:SKL}. Note, however, that the estimation of the virtual $X$-basis QBER for single-photon events, $\phi_{X,1}$, must still be performed using the average QBER, as in the standard block-wise Decoy-State analysis.

\section{Link Budget Model}
\label{sec:sat_model} 
In this work, the considered communication channel is a \ac{FSO} downlink from a \ac{LEO} satellite to an \ac{OGS}. Specifically, we consider a Sun-synchronous orbit, which guarantees that there are nighttime passes every day throughout the year. We fix a link reference scenario which reproduces the KIODO measurement campaign carried out by \ac{DLR}~\cite{kiodo}. The satellite is in a circular Sun-synchronous orbit at altitude $h_\text{sat} = \SI{567}{\kilo\meter}$ above the Earth (using as mean radius $R_E = \SI{6371}{\kilo\meter}$) and consider a pass where the satellite reaches a maximum elevation of $\SI{80}{\degree}$, while the link is established above a minimum elevation of $\SI{20}{\degree}$. For the receiver we employ the physical parameters of the \ac{OGS-KN}. A summary of the parameters employed in the simulations are reported in Table~\ref{constants}, which are realistic for near-term \ac{Sat-QKD} implementations.

\begin{table}[t!]
\setlength{\tabcolsep}{3pt} 
\centering
\begin{tabular}{|l|r|l|l|}
\hline
\textbf{Name} & \textbf{Value} & \textbf{Unit} & \textbf{Description} \\ \hline
\multicolumn{4}{|l|}{\textit{Global System Parameters}} \\ \hline
$\lambda$ & 1550 & nm & Wavelength of the quantum signal \\ \hline
$R_\text{Tx}$ & 1 & GHz & Transmitter pulse repetition rate \\ \hline 
$h_\text{sat}$ & 567 & km & Satellite orbital altitude \\ \hline
$\theta_\text{div}$ & 5 & \si{\micro\radian} & Beam half-divergence angle \\ \hline
$T_\text{pass}$ & 330 & s & Satellite pass duration \\ \hline
$\theta_\text{min}$ & 20 & deg & Minimum elevation angle \\ \hline
$\theta_\text{max}$ & 80 & deg & Maximum elevation angle \\ \hline
$h_\text{OGS}$ & 602 & m & Altitude of the \acl{OGS} \\ \hline
$D_\text{OGS}$ & 0.8 & m & OGS primary mirror diameter \\ \hline
$D_\text{OGS}^\text{int}$ & 0.3 & m & OGS secondary mirror (obscuration) diameter \\ \hline
\multicolumn{4}{|l|}{\textit{Link Budget Parameters}} \\ \hline
$\eta_\text{coll}$ & [sim.] & dB & Receiver collection efficiency \\ \hline
$\eta_\text{atm}$ & [sim.] & dB & Loss from atmospheric absorption \\ \hline
$\eta_\text{scint}$ & [sim.] & dB & Scintillation due to atmospheric turbulence \\ \hline
$\eta_\text{pointing}$ & -3.0 & dB & Loss due to pointing inaccuracies \\ \hline
$\eta_\text{Rx}$ & -1.5 & dB & Internal loss from receiver optics \\ \hline
$\eta_\text{det}$ & -7 & dB & Detector efficiency (equivalent to 20\%) \\ \hline
\multicolumn{4}{|l|}{\textit{Scintillation Simulation Parameters}} \\ \hline
$R_\text{sample}$ & 40 & kHz & Sampling frequency for scintillation simulation \\ \hline
$\tau_\text{corr}^\text{max}$ & 0.03 & s & Time length of the simulation filter \\ \hline
$\Delta$ & 1 & s & Time between simulation filter updates \\ \hline
\multicolumn{4}{|l|}{\textit{QKD Protocol Parameters}} \\ \hline
$p_Z$ & 0.85 & - & Probability of choosing the key generation basis \\ \hline
$\mu$ & 0.59 & - & Mean photon number of the signal state \\ \hline
$\nu$ & 0.21 & - & Mean photon number of the decoy state \\ \hline
$p_\mu$ & 0.80 & - & Probability of sending the signal state $\mu$ \\ \hline
$p_\nu$ & 0.14 & - & Probability of sending the decoy state $\nu$ \\ \hline
$\delta_\text{mis}$ & 5 & deg & Misalignment between Alice and Bob's bases \\ \hline
$R_\text{noise}$ & 3 & kHz & Effective noise photon rate \\ \hline
$t_\text{holdoff}$ & 100 & ns & Detector hold-off (dead) time \\ \hline
$\varepsilon_\text{cor}$ & $10^{-15}$ & - & Correctness parameter of the \ac{QKD} protocol \\ \hline
$\varepsilon_\text{sec}$ & $10^{-9}$ & - & Security parameter of the \ac{QKD} protocol \\ \hline 
\end{tabular} \vspace{2mm}
\caption{Reference parameters for the simulations. }
\label{constants}
\end{table}

In our context, the link budget is a model of the end-to-end efficiency for the \ac{FSO} channel. The factor that we take into account in the end-to-end loss are
\begin{align}
    \eta = \eta_\text{coll} \cdot \eta_\text{pointing} \cdot \eta_\text{atm} \cdot \eta_\text{Rx} \cdot \eta_\text{det} \cdot \eta_\text{scint}
\end{align}
which denote the collection efficiency, pointing loss, atmospheric loss, receiver internal loss, detector efficiency and the (stochastic) scintillation effect.
Since these factors enter multiplicatively in the link budget, they are usually expressed in dB units and, for convenience, we introduce the function
\begin{align}
    \dB[x] := 10 \log_{10} (x) .
\end{align}

\paragraph{Collection efficiency}
The most prominent factor in the link budget is the receiver collection efficiency $\eta$. For communications between terminals that are in the far field from each other that is (approximately) given by the Friis equation~\cite{friis}:
\begin{align}
\label{eta_collection}
    & \eta_\text{coll} = G_\text{Tx} \cdot G_\text{Rx} \cdot L_\text{free space} 
\end{align}
where $G_\text{Tx}$ $(G_\text{Rx})$ is the transmitter (receiver) ideal antenna gain, $L_\text{free space} = \left( \frac{\lambda}{4\pi L} \right)^2$ is the free space loss, 
$L$ is the link distance and $\lambda$ is the wavelength. We assume $\lambda = \SI{1550}{\nano\meter}$, as optical communication technology is most developed for the C-band. 
Upper bounds to the antenna gains are given by $G_\text{Tx} \leq \frac{4\pi A_\text{Tx}}{\lambda^2}$ ($G_\text{Rx} \leq \frac{4\pi A_\text{Rx}}{\lambda^2}$), where $A_\text{Tx}$ $(A_\text{Rx})$ is the transmitter (receiver) antenna aperture area. 
The interpretation of Eq.~\eqref{eta_collection}  is that only a small fraction of the sent signal power is received because the beam spot at the \ac{OGS} is much larger than the diameter of the telescope. Notice that $\eta$ decreases quadratically with the distance $L$. 
The transmitted signal is modeled as a Gaussian beam with half-divergence $\theta_\text{div}$. This is connected to the size of the transmitter aperture, $D_\text{Tx}$, which dictates the minimum beam divergence allowed by diffraction ($\theta_\text{div} \gtrsim \lambda/D_\text{Tx}$). Furthermore, we assume that the receiver telescope is a Cassegrain-type telescope, having a secondary circular mirror of diameter $D_\text{OGS}^\text{int}$ obscuring the primary one, having diameter $D_\text{OGS}$. The gains are thus computed as
\begin{align}
    & G_{\text{Tx}} = \frac{8}{\theta_{\text{div}}^2} 
    ~~~\text{and}~~~ 
    G_{\text{Rx}} = \frac{\pi^2}{\lambda^2} \big(D_\text{OGS}^2 - (D_\text{OGS}^\text{int})^2 \big).
\end{align}
In our model we use $\theta_\text{div}= \SI{5}{\micro\radian}$ (half-divergence to $1/e^2$ intensity angle), which is achievable with current technology~\cite{micius} and reasonably sized aperture on the satellite terminal ($D_\text{Tx} \approx \SI{200}{\milli\meter}$). For the ground telescope we employ the values $D_\text{OGS}=\SI{800}{\milli\meter}$ and $D_\text{OGS}^\text{int}=\SI{300}{\milli\meter}$, which are the physical dimensions of the \ac{OGS-KN}.

\paragraph{Pointing loss}
The small divergence of \ac{FSO} beams requires a correspondingly accurate pointing of the transmitter terminal, as platform vibrations and other pointing inaccuracies result in a decrease of the received signal. The pointing loss, as a function of a stochastic angular jitter with standard deviation $\sigma_\text{jitter}$ and systematic error $\theta_\text{bias}$ is given by~\cite{carrillo_2023}
\begin{align}
    \eta_\text{pointing} = 
    \frac{\theta_\text{div}^2}{\theta_\text{div}^2 + 4\sigma_\text{jitter}^2} \exp\!\left(-\frac{2\theta_\text{bias}^2}{\theta_\text{div}^2 + 4\sigma_\text{jitter}^2}\right) .
\end{align}
However, in a closed loop system there is a complex dependence of the jitter on the system parameters (including on the beam divergence itself) which is difficult to model. While a pointing jitter as small as $\sigma_\text{jitter}=\SI{0.47}{\micro\radian}$ has been demonstrated on a satellite downlink~\cite{wang_2021}, which for a beam half-divergence $\theta_\text{div}=\SI{5}{\micro\radian}$ would correspond to a loss of only $-0.15\,\text{dB}$, we prefer to use a more conservative estimate of $\sigma_\text{jitter}=\SI{2.5}{\micro\radian}$, resulting in a pointing loss $\dB[\eta_\text{pointing}] = -3.0$.

\paragraph{Atmospheric loss}
The propagation of an optical signal through the atmosphere results in losses due to absorption and scattering. The attenuation is described by a function $\alpha(h, \lambda)$ of the altitude $h$ and of the wavelength $\lambda$ and is typically expressed in dB/km. Then, the atmospheric loss can be computed as
\begin{align}
   \dB[\eta_{\text{atm}}] = \frac{\int_{h_\text{OGS}}^{h_\text{sat}} \alpha(h, \lambda) \, dh}{\cos({\zeta})} 
\end{align}
where $\zeta$ is the zenith angle of the satellite as seen at the OGS. Furthermore, $\alpha(h, \lambda)$ varies depending on location and weather conditions. We have fitted $\alpha(h, \lambda)$ from MODTRAN simulations~\cite{modtran}, assuming a ground level visibility of $\SI{23}{\kilo\meter}$, which represents an average working visibility condition. We have set $h_\text{OGS} = \SI{602}{\meter}$, the altitude above sea level of the OGS-KN.

\paragraph{Receiver internal loss}
The receiver internal loss describes the attenuation within the receiver terminal, e.g., due to power attenuation on the optical surfaces. This loss depends on the total number of surfaces on the path of the optical signal and on their quality. Here we assume $7$ mirrors with $95\%$ reflectivity each, giving a total loss of around $\SI{-1.5}{dB}$.

\paragraph{Transmitter internal loss}
For single-photon sources, transmitter internal losses must be taken into account as they directly impact the number of transmitted photons. However, in the case of Decoy-State \ac{BB84} with attenuated laser sources such losses can often be neglected. This is because the transmitted signal is already strongly attenuated to reach the few-quanta regime and set the pulse intensity used in the Decoy-State method. Therefore, if the value of the transmitter internal loss is known, it can be precompensated by including it in the total signal attenuation.

\paragraph{Detector efficiency}
A critical aspect is the final coupling of the signal to the detector, which presents a non-trivial trade-off between two solutions: the use of fiber-coupled detectors of free-space-coupled ones. The analysis of the coupling efficiency from free-space to fiber is rather complex and depending on an accurate model of the employed \ac{AO} system. For ease of analysis, in this work we have modeled the detector a free-space InGaAs \ac{SPAD}. This is a technology suitable for operation in the C-band, having an efficiency of $20\%$, approximately equal to $\SI{-7}{dB}$ of loss, and a hold-off time after each detection of $\SI{100}{\nano\second}$.

\section{Model of Atmospheric Scintillation}
\label{sec:scintillation}

The propagation of an optical signal through atmospheric turbulence results in intensity scintillation of the beam. The origin of scintillation are wave-front distortions, resulting in self-interference of the beam after a certain propagation distance. A parameter that is used to evaluate the power fluctuations observed by the receiver due to scintillation is the \ac{PSI}
\begin{align}
    \sigma_\text{scint}^2 = \langle \eta_\text{scint}^2(t) \rangle - 1
\end{align}
where we assume the normalization $\langle \eta_\text{scint}(t) \rangle = 1$. The fluctuation in the received optical power (proportional to $\eta_\text{scint}$) is averaged over a time that is longer than time scale over which the scintillation varies, given by the inverse of the Greenwood's frequency. 

\subsection{Modified Hufnagel-Valley Turbulence Model}
Turbulence causes small-scale fluctuations in the index of refraction in Earth's atmosphere with a spatial correlation that, according to Kolmogorov's theory, approximately follows a power law. These fluctuations are primarily due to random variations in temperature and pressure, which lead to inhomogeneities in air density and, ultimately, to refractive index variations. As an optical beam propagates through these turbulent layers, these refractive index variations distort the phase front and redistribute the beam intensity, resulting in intensity scintillation.

The turbulence strength is quantified by the so-called index of refraction structure constant, $C_n^2$. Its dependence on the altitude above seal level, $h$, is modeled starting from \ac{HV} atmospheric turbulence profile~\cite{hufnagel,valley}, which is the de-facto standard profile for modeling of atmospheric turbulence~\cite{field_guide}. We employ a modified \ac{HV} model, following the work in ~\cite{correction_factor}, which includes a correction factor $F$ for the strength of turbulence in the boundary layer when the terrain is at a certain altitude $h_\text{OGS}$ above sea level. This is given by
\begin{align}
    C_n^2(h) &= 
    0.00594 \left(\frac{w}{27}\right)^2 (h \cdot 10^{-5})^{10} e^{-h/1000} + \nonumber\\
    & + 2.7 \times 10^{-16} e^{-h/1500} + A F^2 e^{-\frac{(h - h_{\text{OGS}})}{100}}
\end{align}
where $w = \SI{21}{\meter/\second}$ is the effective wind speed, $A = \SI{1.7e-14} {\meter^{-2/3}}$ is the turbulence at ground level. The correction factor 
is $F=0.93$ for $h_\text{OGS}=\SI{602}{\meter}$.

\subsection{Power Scintillation Index Calculation}

The calculation of the \ac{PSI} requires several intermediate steps. First, we start from computing Rytov parameter $\sigma_\text{Ry}^2$, which corresponds to the (normalized) variance of the received power for point-like receivers, i.e.\ considering the optical intensity fluctuation on the axis of the beam, in the so-called weak turbulence condition. According to Rytov's first order perturbation theory, for a downlink beam (for which the plane wave approximation also holds) this quantity is given by
\begin{align}
    \sigma_\text{Ry} &= \frac{2.25k^{7/6}}{\cos^{11/6} \zeta}
    \int_{h_\text{OGS}}^{h_\text{sat}} C_n^2(h) (h - h_{\text{OGS}})^{5/6} dh
\end{align}
where $k = 2\pi/\lambda$ is the optical wave number and $\zeta$ is the zenith angle of observation.

Second, we need to take into account aperture averaging, using the approach presented in~\cite{yura}. A beneficial effect in \ac{FSO} is that an extended receiver observes significantly reduced scintillation compared to a point-like receiver, especially when the beam spatial coherence length is much smaller than telescope aperture. This stems from the fact that adjacent regions of high and low power intensity within the telescope aperture will partially average each other out. First, we need to compute an effective link length, related to the length of the section of the link which is significantly affected by turbulence
\begin{align}
    L_\text{eff} =  \frac{1}{\cos\zeta}\left(\frac{18 \int C_n^2(h) h^2 dh}{11 \int C_n^2(h) h^{5/6} dh} \right)^{\!6/7} 
\end{align}
from which we can derive the aperture averaging scaling factor
\begin{align}
    f_\text{aa} = \left(1+1.062d_\text{aa}^2\right)^{-7/6}, \qquad
    d_\text{aa} = D_\text{OGS} \sqrt{\frac{k}{4L_\text{eff}}}
\end{align}
so that the \ac{PSI} for an extended receiver is $\sigma_\text{scint}^2 = f_\text{aa}  \sigma_\text{Ry}^2$. However, this is valid only in the weak turbulence regime, which is not applicable when the satellite is at a low elevation angle. In order to derive an expression that applies to both the weak and strong turbulence conditions, we employ the following expression for the PSI, stemming from the use of the so-called extended Rytov theory~\cite[Chapter~10.3.2]{laser-prop}:
\begin{align}
    \sigma_\text{scint}^2 = 
    \exp\!\Bigg[ & \frac{0.49 \sigma_\text{Ry}^2}{(1 + 0.65 d_{\text{aa}}^2 + 1.11 \sigma_\text{Ry}^{12/10})^{7/6}} + \nonumber\\ 
    +& \frac{0.51 \sigma_\text{Ry}^2 (1 + 0.69 \sigma_\text{Ry}^{12/10})^{-5/6}}{1 + 0.9 d_{\text{aa}}^2 + 0.62 d_{\text{aa}}^2 \sigma_\text{Ry}^{12/10}} \Bigg] - 1
\end{align}

\subsection{Greenwood's Frequency}

The inverse of Greenwood's frequency defines the time scale over which the atmospheric turbulence varies. It is determined by the velocity with which turbulence cells move in front of the field of view of the telescope. This depends both on the wind speed and on the satellite motion. In this work we consider low wind speed conditions, so that we only need to consider the dominant contribution arising from the slew rate of the line-of-sight, for which we now derive an approximate expression. 

From the orbit propagation software one can get the component of the satellite velocity orthogonal to the line-of-sight $v_\text{sat}^\perp$. 
From this a virtual wind speed associated to the slew rate of the line-of-sight can be computed as
\begin{align}
    V(h) = \frac{h-h_\text{OGS}}{h_\text{sat}-h_\text{OGS}} v_{\text{sat}}^\perp .
    \label{eq:pseudowind}
\end{align}

From the pseudo-wind $V(h)$ and the turbulence profile $C_n^2(h)$ one can compute Greenwood's frequency
\begin{align}
    f_\text{G} = 2.31 \lambda^{-6/5} \left(\frac{1}{\cos \zeta}\int C_n^2(h) V(h)^{5/3} dh\right)^{\!3/5} .
    \label{eq:greenwood}
\end{align}

\subsection{Scintillation Signal}
\label{sec:signal}

The \ac{PDF} of the scintillation signal $\eta_\text{scint}$ in a \ac{FSO} downlink is accurately modeled by a lognormal distribution~\cite{laser-prop} and, furthermore, the \ac{PDS} is well approximated by a fourth order Butterworth low-pass filter~\cite{butterworth,latorre}. We thus model the \ac{PDS} of the scintillation signal in the Laplace domain as
\begin{align}
    H_{(t)}(s) = \left(1+\left(\frac{s}{2\pi f_\text{G}(t)}\right)^{2n}\right)^{\!-1/2}
    \label{eq:filter}
\end{align}
where $n=4$ is the order of the filter and the cut-off frequency is given by Greenwood's frequency $f_\text{G}(t)$ which slowly varies in time during the satellite pass. The \ac{PDF} of the scintillation signal is lognormal, given by  
\begin{align}
    p_{(t)}(\eta_\text{scint})
    =
    \frac{1}{\eta_\text{scint}\sqrt{2\pi\Sigma^2(t)}} \exp\!\left(-\frac{\left[\ln(\eta_\text{scint})+\frac{1}{2}\Sigma^2(t)\right]^{2}}{2\Sigma^2(t)}\right)
    \label{eq:lognormal}
\end{align}
with $\Sigma^2(t) = \ln(\sigma_\text{scint}^{2}(t)+1)$. This \ac{PDF} has mean 1 and has variance $\sigma_\text{scint}^2(t)$ which slowly changes in time.

\section{Model of the QKD Signal}
\label{sec:QKD_model}

Here we model the signal observed by the receiver in a Decoy-State BB84 protocol with active basis choice~\cite{hausler}. The receiver consists of a polarization rotation stage, which may apply a $\SI{45}{\degree}$ rotation to the incoming photons, followed by a polarizing beam splitter that directs the signal to two single-photon detectors. The detectors are modeled as ideal threshold devices, registering a click whenever one or more photons arrive. We label them  as correct ($\checkmark$) and incorrect ($\mathcal{X}$), depending on whether the outcome matches Alice's transmitted polarization. 

When Alice transmits pulses at a rate $R_\text{Tx}=\SI{1}{\giga\hertz}$ with mean photon number $\alpha$ (signal $\alpha=\mu$, decoy $\alpha=\nu$, or vacuum $\alpha=0$), the expected number of received photons is $\eta(t) \alpha$, where $\eta$ includes  channel transmittance and detector efficiency. The total expected photon numbers at the detectors are 
\begin{align}
\begin{cases}
    \tau_{\ok,\alpha} = \eta(t) \cdot \alpha \cos^2(\delta_\text{mis}) + r_\text{noise}, \\
    \tau_{\ko,\alpha} = \eta(t) \cdot \alpha \sin^2(\delta_\text{mis}) + r_\text{noise},    
\end{cases}
\end{align}
where  $\delta_\text{mis}=\SI{5}{\degree}$ is the misalignment between the transmitter and receiver reference frames, and  $r_\text{noise}$ accounts for background light and detector dark counts. The noise term is expressed as the number of effective noise photons per detection window of duration $1/R_\text{Tx}$, i.e.,  $r_\text{noise} = R_\text{noise}/R_\text{Tx} = 3\cdot 10^{-6}$ for a noise photon rate $R_\text{noise} = \SI{3}{\kilo\hertz}$. 

Since photon arrivals follow a Poisson distribution, the click probabilities for the two detectors are
\begin{equation}
    p(\ok|\alpha) = 1 - e^{-\tau_{\ok,\alpha}}, \quad p(\ko|\alpha) = 1 - e^{-\tau_{\ko,\alpha}}.
\end{equation}
If both detectors click, the outcome is randomly assigned to one of them.

The resulting \ac{QBER} is the ratio of erroneous to total  detection events (including both  single-clicks and double-clicks):
\begin{align}
\label{eq:qber}
    Q(\alpha) = 
    \frac{p(\text{err}|\alpha)}{p(\text{click}|\alpha)} =
    \frac{p(\ko|\alpha) - \frac{1}{2} p(\ok|\alpha) p(\ko|\alpha)}{p(\ok|\alpha) + p(\ko|\alpha) - p(\ok|\alpha) p(\ko|\alpha)} .
\end{align}
This expression applies to both the $Z$ and $X$ bases. Note that $Q(\alpha)$ decreases monotonically with $\alpha$ and satisfies $Q(0)=\tfrac{1}{2}$.

\section{Simulation Methodology}
\label{sec:methodology}

\begin{figure}[t!] 
    \centering
    \includegraphics[width=\columnwidth]{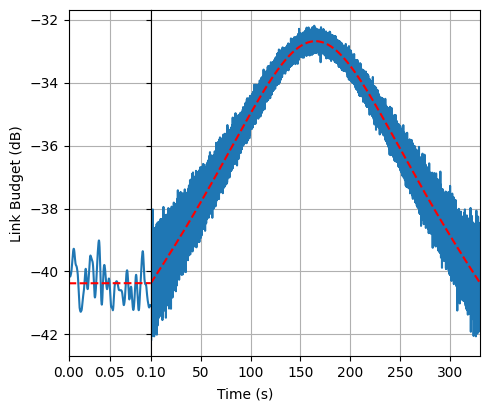} \vspace{-6mm}
    \caption{Simulated downlink end-to-end efficiency (i.e., link budget) for the reference satellite pass, including (solid, blue) and excluding (dashed, red) scintillation effects. Left: zoom-in of the first $\SI{100}{\milli\second}$ of the pass.} 
    \label{fig:scintillation} 
\end{figure}

The satellite orbit was selected from the KIODO measurement campaign conducted by \ac{DLR}~\cite{kiodo}. The simulated pass has a total duration of $T_\text{pass}=\SI{330}{\second}$. For each second, the corresponding elevation angle and link distance are provided by the orbit propagation software. 

The simulation is structured in three main parts: (i)  generation of a scintillation time series, (ii)  simulation of \ac{QKD} signal transmission over the resulting channel, and (iii) execution of an \ac{IR} protocol on the simulated data.

\subsection{Generation of the Scintillation Time Series}

In order to generate a simulated scintillation signal with a specified \ac{PDF} and \ac{PDS}, we employ the approach introduced in~\cite{gujar}. The procedure starts  by drawing normalized Gaussian random variables $\textbf{u}=\{u(t_i)\}_i$ where the times $t_i$ are sampled at a sufficiently high frequency, here chosen as $R_\text{sample} = \SI{40}{\kilo\hertz}$. A linear filter $\textbf{h}$ is then applied to the vector $\textbf{u}$ to introduce the time correlations specified by the \ac{PDS}, giving $\textbf{v} = \textbf{h} * \textbf{u}$. This is followed by a zero-memory (i.e., point-wise) nonlinear transformation to match the target \ac{PDF}. 

The slowly-varying Butterworth filter in~\eqref{eq:filter} is applied to the array $\textbf{u}$ of Gaussian random variables. Typically, the filter is applied in the Fourier domain, where convolution reduces to point-wise multiplication and is thus computationally efficient. However, due to the explicit time dependence of the filter, we instead perform the operation in the time domain by convolving with $h_{(t)}(\tau)=\mathcal{L}^{-1}\{H_{(t)}\}(\tau)$, the inverse Laplace transform of~\eqref{eq:filter}. For computational efficiency, we introduce a parameter $\tau_\text{corr}^\text{max}$ beyond which correlations among scintillation values are considered negligible, and approximate $h_{(t)}(\tau) \simeq 0$ for $\tau> \tau_\text{corr}^\text{max}$. Here we set $\tau_\text{corr}^\text{max} = \SI{0.03}{\second}$, such that  $h_{(t)}(\tau)$ is evaluated at $n_\text{max}=R_\text{sample} \tau_\text{corr}^\text{max} = 1200$ points. We then define the function
\begin{align}
    \widetilde{h}_{(t)}(\tau) = 
    \begin{cases}
        h_{(t_\Delta)}(\tau) & \text{if } \tau \leq \tau_\text{corr}^\text{max}, \\
        0 & \text{otherwise},
    \end{cases}
\end{align}
where $t_\Delta = \lfloor t/\Delta +0.5\rfloor \Delta$, meaning that $\widetilde{h}_{(t)}$ is updated once per time interval $\Delta$. In this work we set $\Delta=\SI{1}{\second}$. The filtering operation is then expressed as the discrete convolution
\begin{align}
    v(t_i) = (\widetilde{h}_{(t_i)} * u)(t_i) = \sum_{j=0}^{n_\text{max}} \widetilde{h}_{(t_i)}(t_j)u(t_i-t_j).
\end{align}
The overall time complexity of the algorithm is $O(n_\text{max} N_\text{tot})$, where the total number of points is $N_\text{tot} = T_\text{pass} R_\text{sample} = 13.2\cdot 10^6$.

Next, the zero-memory nonlinear transformation $\mathcal{F}_{(t)}(V) = \exp(V \Sigma(t) -\Sigma^2(t)/2)$ is applied, mapping Gaussian variables to the lognormal distribution $p_{(t)}$ defined in~\eqref{eq:lognormal}. The resulting scintillation signal is $\eta_\text{scint}(t) = \mathcal{F}_{(t)}(v(t))$.

Finally, the scintillation signal is combined with the rest of the link budget to obtain the end-to-end efficiency $\eta(t)$ over the entire satellite pass. The result is shown in Fig.~\ref{fig:scintillation}.

\subsection{Generation of Simulated QKD Signals}

The global transmission $\eta(t)$ is then used to simulate a Decoy-State BB84 protocol. The outputs consist of an array indicating the click times at Bob's receiver and, for those corresponding to $Z$-basis measurements (the basis used for key generation), Alice's sifted key $\mathbf{z}_A$ and the error vector $\mathbf{e}$, where Bob's sifted key is $\mathbf{z}_B = \mathbf{z}_A \oplus \mathbf{e}$. In addition, an array containing the \ac{QBER} associated with each click is generated. This  information is crucial for enhancing the performance of the \ac{IR} procedure. 

We first generate the click instants $\textbf{c} = \{c_i\}_{i=1}^{n_\text{click}}$.  The possible click instants are expressed as integers between $0$ and $T_\text{pass}R_\text{Tx}-1$, where $T_\text{pass}R_\text{Tx} = 330\cdot 10^{9}$ is the total number of transmitted pulses. Since the probability of a click is very low (always less than $10^{-3}$ in our scenario), one could, in principle, use a geometric random variable $G \sim \text{Geo}(p_{\text{click}})$ to directly sample the next click instant, where $p_{\text{click}}$ is obtained by averaging $p(\text{click}|\alpha)$ over $\alpha$.

However, two aspects must be taken into account: (i) the detector remains inactive for a time $t_{\text{hold-off}}$ after each click, and (ii)  $p_{\text{click}}$ depends on $\eta(t)$ and thus varies with time. Since the downlink transmission $\eta(t)$ is sampled at a rate $R_\text{sample}$,  $p_{\text{click}}$ only changes with this frequency. The number of transmission instants for which the conditions remain constant is therefore
\begin{align}
    n_\text{window} = \frac{R_\text{Tx}}{R_\text{sample}}.
\end{align}

Let $c_i$ denote the current click instant. The next click instant $c_{i+1}$ is sampled as follows. We initialize an auxiliary variable $\theta\leftarrow c_i+n_\text{hold-off}$, where $n_{\text{hold-off}}=t_{\text{hold-off}} \cdot R_\text{Tx}$ is the number of instants during which the detector is inactive after a click. We then draw  $G\sim\text{Geo}(p_\text{click}(\theta))$ and set a tentative time increment $\theta' \leftarrow  \theta+G$. If (case $a$) the condition  $\lfloor\theta'/n_\text{window}\rfloor = \lfloor\theta/n_\text{window}\rfloor$ holds, then $\theta'$ lies within an interval in which $p_{\text{click}}$ is constant, and we set $c_{i+1} = \theta'$. Otherwise (case $b$) $\theta'$ exceeds the interval, making the draw invalid. In this case, by exploiting the memoryless property of the process, we advance $\theta$ to the start of the next window, i.e.,  $\theta\leftarrow n_\text{window} (\lfloor\theta/n_\text{window}\rfloor+1)$, and repeat the procedure until case $a$ is satisfied. This process continues until the end of the satellite pass, yielding the vector of click instants $\mathbf{c}$.

Each click is then assigned, a posteriori, the bases in which Alice prepared the photon and Bob performed the measurement, with probabilities $p_Z$ and $p_X=1-p_Z$, so that the joint probability of both using the $Z$ basis is $p_Z^2$. For each $Z$ event, a uniformly random value is drawn to generate Alice's sifted key $\textbf{z}_A$. Next, an intensity $\alpha$ is assigned to each click event according to $p(\alpha|\text{click})$, obtained via Bayes' rule from the known values $p(\text{click}|\alpha)$ and $p_\alpha$. A vector of QBER values $\textbf{q} =\{q_i\}_i$ is then computed for each sifted $Z$ bit using Eq.~\eqref{eq:qber}. Finally, the error vector $\textbf{e}$ is constructed by setting each bit $e_i$ to $1$ with probability $q_i$ and to $0$ otherwise.

\subsection{Simulation of IR}

Finally, \ac{IR} is performed on the simulated data using a \ac{DLL} that implements the rateless protograph \ac{LDPC} codes described in~\cite{protograph}, kindly provided by the authors. The inputs to the library are: $\Phi$, a scalar \ac{QBER} parameter used to select the initial code family; $\mathbf{z}_A$ and $\mathbf{z}_B$, the sifted keys of Alice and Bob; and an array of instantaneous  \ac{LLR} values $\mathbf{llr}$. The \ac{LLR} values convey the reliability of each bit to the decoder and are computed as
\begin{align}
    \text{llr}_i = \ln\!\left(\frac{1-q_i}{q_i}\right).
\end{align}

The results of the \ac{IR} simulations, obtained under different strategies to identify the most effective configurations, are presented in Sec.~\ref{sec:simulations}.

\section{Simulation results}
\label{sec:simulations}

\begin{figure*}[t!]
    \centering
    \setlength{\tabcolsep}{0pt}

    \begin{tabular}{ccc}
        \subcaptionbox{\label{fig:Combined:a}}{\includegraphics[width=0.34\textwidth]{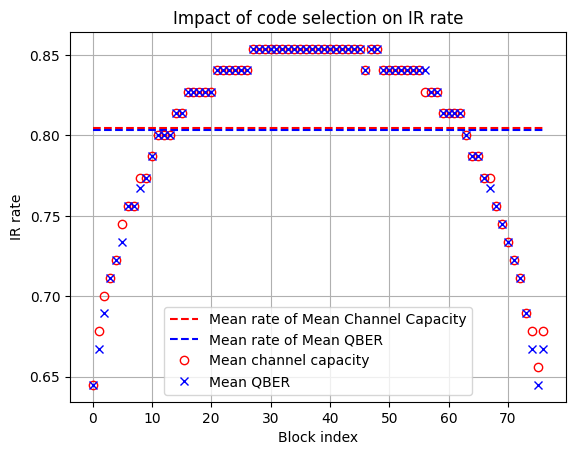}} &
        \subcaptionbox{\label{fig:Combined:b}}{\includegraphics[width=0.34\textwidth]{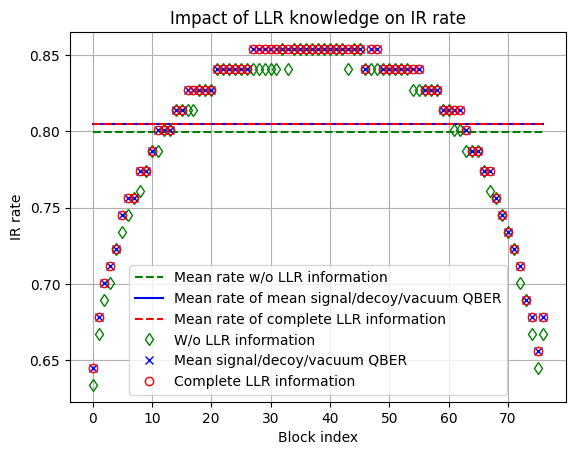}} &
        \subcaptionbox{\label{fig:Combined:c}}{\includegraphics[width=0.34\textwidth]{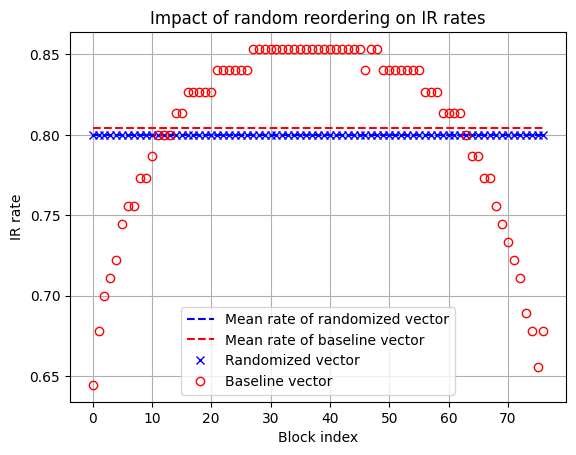}} 
        \\
        \subcaptionbox{\label{fig:Combined:d}}{\includegraphics[width=0.34\textwidth]{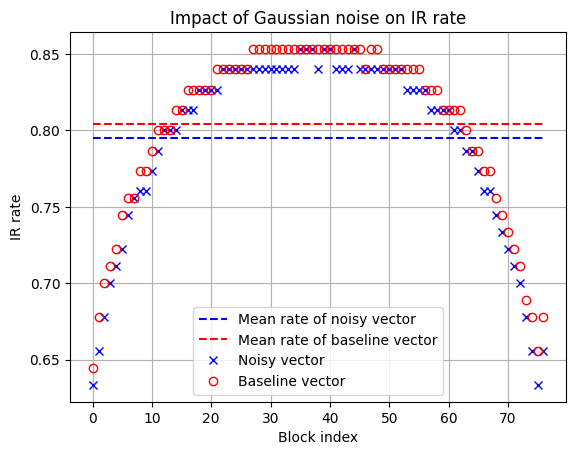}} &
        \subcaptionbox{\label{fig:Combined:e}}{\includegraphics[width=0.34\textwidth]{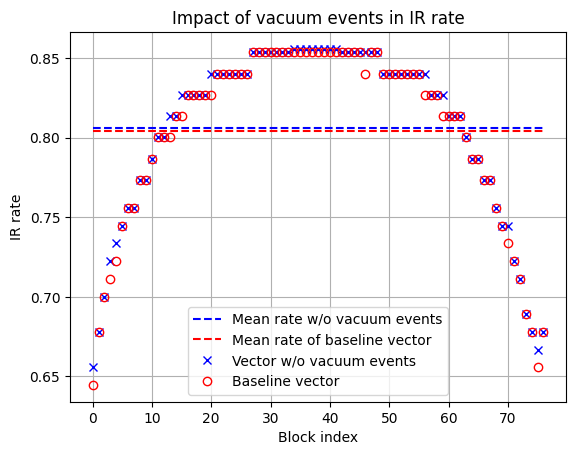}} &
        \subcaptionbox{\label{fig:Combined:f}}{\includegraphics[width=0.34\textwidth]{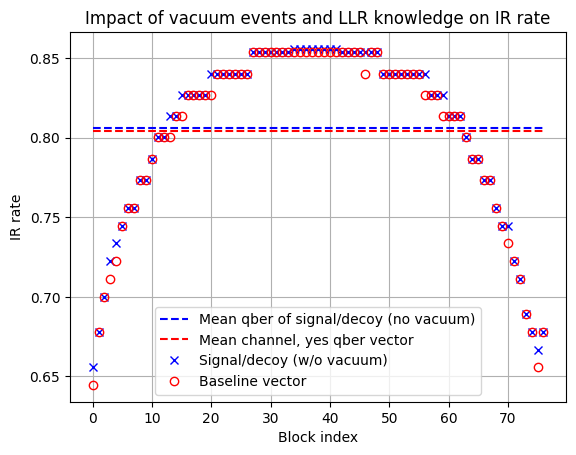}}
    \end{tabular}

    \caption{
    On the $x$-axis are the indexes of the blocks used in \ac{IR}, and on the $y$-axis are the corresponding \ac{IR} rates, obtained via Eq.~\eqref{eq:rate} from the \ac{DLL} output.
    (a) Mean channel capacity vs.\ mean QBER for code selection; in the following figures the mean channel capacity is always used for code selection. 
    (b) Complete LLR knowledge vs.\ mean signal/decoy/vacuum vs.\ mean block-wise QBER; in the following figures complete LLR knowledge is used as baseline. 
    (c) Baseline vs.\ bit position randomization strategy. 
    (d) Baseline vs.\ noisy LLR information. 
    (e) Baseline vs.\ complete LLR without vacuum states.
    (f) Baseline vs.\ mean signal/decoy without vacuum states.
    }
\end{figure*}

The \ac{QBER} array obtained from the \ac{QKD} simulation contains more than $35$ million values. For \ac{IR}, the sequence is divided into blocks of length $N_\text{bl}=460800$, which is a multiple of $1800$, as required by the codes in~\cite{protograph}. The last remaining $k$ bits are assigned to a block of length equal to the largest multiple of $1800$ not exceeding $k$, to minimize discarded bits. Moreover, as shown in Fig.~\ref{fig:scintillation}, the signal tails correspond to the lowest transmittance and thus the highest \ac{QBER} values, leading to poorer reconciliation rates.

The first comparison addresses the choice of $\Phi$ for code selection: either the mean \ac{QBER} in the block or the \ac{QBER} corresponding to the mean channel capacity of the block. In~\cite{protograph}, this distinction is not considered because a constant \ac{QBER} is assumed, and the two coincide. Although the former is a simpler approximation, the latter provides a code rate matched to the effective information transmitted through the noisy channels.

With the mean \ac{QBER} approach,
\begin{align}
    \Phi_j = \frac{1}{N_\text{bl}} \overset{N_j+N_\text{bl}-1}{\underset{i=N_j}{\sum}}q_i,
\end{align}
where $N_j$ is the starting index of the $j$-th block. Alternatively, each bit is modeled as being transmitted through its own \ac{BSC} with error probability $q_i$. The mean capacity of the sequence of \ac{BSC}s is then
\begin{align}
    C_j = 1-\frac{1}{N_\text{bl}}\overset{N_j+N_\text{bl}-1}{\underset{i=N_j}{\sum}}h_2(q_i), 
\end{align}
and the equivalent \ac{QBER} $\Phi_j$  is obtained by solving  $C_j=1-h_2(\Phi_j)$ by bisection. 

As shown in Fig.~\ref{fig:Combined:a}, the red curve (mean channel capacity) is consistently above the blue curve (mean QBER), except for block~$56$, likely due to  statistical fluctuations. The average rate is also higher. Therefore, all subsequent simulations adopt $\Phi$ computed via the mean channel capacity.

A second study evaluates the impact of different strategies for passing the optional variable $\mathbf{llr}$: 
\begin{enumerate}
    \item Excluding $\mathbf{llr}$ entirely.
    \item Using block-wise average \ac{QBER} values for each intensity (signal, decoy, vacuum). In other words, if a particular instant clicked when the state was the decoy, the $\mathbf{llr}$ array in that particular instant will have as a value the average \ac{QBER} of decoy states for the entire block (and analogously for the signal and vacuum states).
    \item Using the full QBER array from the \ac{QKD} simulation.
\end{enumerate}
Figure~\ref{fig:Combined:b} shows that strategies~$2$ and~$3$ yield identical rates for each block and significantly outperform strategy~$1$. This demonstrates that comparable performance can be achieved either by retaining full \ac{LLR} information or, more efficiently, by tracking only block-wise averages of the signal, decoy, and vacuum states.

For subsequent comparisons, the curve with full \ac{LLR} information (red in Fig.~\ref{fig:Combined:b}) is taken as the baseline. 

A further strategy was to randomize the order of detection events, with the rationale that high-quality signals (central part of the curve) could assist in correcting lower-quality ones (tails). However, Fig.~\ref{fig:Combined:c} shows that the randomized rates, now constant across blocks, are slightly lower than the average  baseline.

Another approach was to corrupt the $\mathbf{llr}$ values  by adding Gaussian noise with mean $0$ and standard deviation $0.125$. As shown in Fig.~\ref{fig:Combined:d}, this significantly reduces the rates, highlighting the importance of accurate \acs{LLR} estimation.

\begin{figure}[t!] 
    \centering
    \includegraphics[width=\columnwidth]{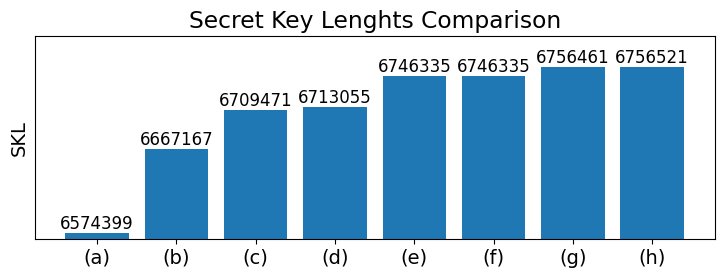} 
    \caption{
    Comparison of the \ac{SKL} obtained under the following strategies:
    a)~no $\mathbf{llr}$; 
    b)~baseline (case (e)) with bit position randomization; 
    c)~noisy LLR estimation; 
    d)~LLR vector with mean QBER for code selection; 
    e)~LLR vector with mean channel capacity for code selection (baseline); 
    f)~three LLR values for signal/decoy/vacuum; 
    g)~two LLR values for signal/decoy, excluding vacuum events;
    h)~baseline (case (e)), excluding vacuum events.
    } 
    \label{fig:SKL} 
\end{figure}

Next,  the \acs{DLL} was provided with a modified \ac{LLR} vector $\textbf{llr}'$, identical to $\textbf{llr}$ but  
excluding vacuum events. Since vacuum states yield \acs{QBER} $\simeq 0.5$, they may require more bits to correct than their contribution to the sifted key. Figure~\ref{fig:Combined:e} shows that this strategy improves the reconciliation rate compared  to the baseline, but since the sifted key length is reduced (by about $90\,000$ vacuum events), the overall \ac{SKL} improvement is marginal. Indeed, as confirmed in Fig.~\ref{fig:SKL}, this approach yields only a slight increase.

Finally, the same vacuum exclusion strategy was combined with block-wise averages for signal and decoy states. As shown in Fig.~\ref{fig:Combined:f}, this performs almost identically to the full $\textbf{llr}$ vector without vacuum events. The latter produces only $60$ additional secret key bits over nearly seven million, confirming the negligible difference. 

\section{Conclusions}
\label{sec:conclusions}

In this work, we improved the performance of information reconciliation (IR) for Decoy-State BB84 in a simulated satellite downlink scenario. To achieve this, we employed a comprehensive channel model that accounts for the time-varying link budget and a realistic quantum receiver model, allowing us to accurately simulate the instantaneous \ac{QBER} throughout a satellite pass. These values were then used to generate log-likelihood ratio information for the decoder, thereby improving efficiency, reducing information leakage to the eavesdropper, and ultimately yielding a longer secure key from the same  quantum communication data.

Our primary finding is that leveraging a-priori knowledge of the instantaneous \ac{QBER} to inform the low-density parity check (LDPC) decoder enhances the efficiency of \ac{IR}, increasing the secret key length (SKL) by nearly 3\% compared to the standard approach that assumes a uniform error rate across the block. Several reconciliation strategies were evaluated to identify the most effective scheme. We found that  initializing rateless code selection based on the mean channel capacity of each block consistently outperforms using the mean \ac{QBER}. Performance was observed to be highly sensitive to the accuracy of  \ac{QBER} estimation. Notably, a simpler strategy that assigns  block-wise average \ac{QBER} values for signal, decoy, and vacuum states performed nearly identically to the full instantaneous-information case, offering a practical trade-off between complexity and performance. The highest \ac{SKL}, albeit by a narrow margin, was obtained by excluding detections from vacuum states; despite producing a smaller sifted key, this approach yielded a slightly longer final secret key. By contrast, strategies such as bit-order randomization degraded performance. A summary of these results in shown in Fig.~\ref{fig:SKL}.

Overall, the \ac{LDPC}-based \ac{IR} scheme demonstrated very high performance compared with alternatives in the literature, with the inefficiency factor ranging from $f=1.11166$ for the best-performing strategy to $f = 1.1386$ for the worst case, where no instantaneous \ac{QBER} information was provided.

Future research may explore alternative bit-ordering methods to further optimize decoder performance. Additionally, investigating the real-time implementation of these schemes in hardware would be a valuable  step toward practical deployment of more efficient satellite \ac{QKD} systems.

\section*{Acknowledgments}
We thank Dr.~Alberto Tarable for adapting the software implementing message passing decoding for protograph LDPC codes. This work was done within the project QuNET funded by the German Federal Ministry of Education and Research under the funding code 16KIS1265.

\bibliographystyle{IEEEtran}

\bibliography{bibliography}
\end{document}